\documentclass[twocolumn,showpacs,preprintnumbers,amsmath,amssymb,pre]{revtex4}

\usepackage{epsfig}
\begin{document}

\title{An out-of-equilibrium model of the distributions of wealth}

\author{Nicola Scafetta$^{1,2}$, Sergio Picozzi$^{1}$  and Bruce J. West$^{1,2,3}$ \\
1)Physics Department, Duke University, Durham, NC 27708\\
2) Pratt School of EE department, Duke University, Durham, NC 27708\\
3) Mathematics Division, US Army Research Office, Research Triangle Park, NC
27709}

\date{\today}

\begin{abstract}
The distribution of wealth among the members of a society is herein assumed
to result from two fundamental mechanisms, trade and investment. An
empirical distribution of wealth shows an abrupt change between the
low-medium range, that may be fitted by a non-monotonic function with an
exponential-like tail such as a Gamma distribution, and the high wealth
range, that is well fitted by a Pareto or inverse power-law function. We
demonstrate that an appropriate trade-investment model, depending on three
adjustable parameters associated with the total wealth of a society, a social
differentiation among agents, and economic volatility referred to as investment can successfully reproduce the distribution of empirical wealth data in  the low, medium and high
ranges. Finally, we provide an economic interpretation of the mechanisms in
the model and, in particular, we discuss the difference between Classical and Neoclassical  theories regarding the concepts of {\it value} and {\it price}. We consider  the importance that out-of-equilibrium trade transactions, where the prices differ from values, have  in real economic societies. \\



\end{abstract}
\maketitle



\section{Introduction}

In this paper we attempt to replicate the empirical  distribution of wealth using a parsimonious model in which there are two ways by which wealth can be accumulated: by investment and by trading. By investment we intend any act that creates or destroys wealth, whereas by trade we intend any type of economic transaction.  While a basic investment mechanism can be modeled with a stochastic multiplicative model \cite{bouchaud}, herein we focus on a  basic nonlinear stochastic trade mechanism that captures the following properties: a) in trades there may be a transfer of wealth from one agent to the other because the price paid fluctuates around an equilibrium price (=value)  and, therefore, the price may differ from the value of the commodity transferred;  b) in a trade
transaction the amount of wealth that may move from one agent to the other is bounded because the price and the value of a commodity cannot (usually) exceed the wealth of the poorer of the two traders; c)  the price is socially determined in such a way that the trade is statistically biased in favor of
the poorer trader. These properties imply an intrinsic nonlinear stochasticity and can not be captured by  linear econophysics models such as  one based on a mean field approximation \cite{bouchaud}, or  on a kinetic theory approximation \cite{dragulesco1}. We refer to our model as 
the nonlinear stochastic trade-investment  model (NSTIM). 

There is a  difference between the price paid for  and the value of the commodity transferred in a trade. This difference is the basis of NSTIM  and  requires an economic explanation. In fact, the concept of a divergence between price and value is foreign to Neoclassical economic theory, which assumes that all  trades occur in ``equilibrium" \cite{foley} where price and value are equal. This economic assumption is summarized in the so-called {\it ``Law of one price"} \cite{lawoneprice}. Neoclassic economists do not expect trade to involve a transfer of wealth, but rather an increase in utility for both parties  with a zero net transfer of wealth. This balance was formalized into the controversial {\it ``Say's Principle"} \cite{say}.  While the concepts of price and value have effectively been melded in Neoclassical theory, its predecessor, the Classical school of thought, maintained a distinction. Value reflected the long-term cost of production of a commodity in Classical economics, whereas price was what was paid in the market on any particular day. The Classical economists expected divergences to occur regularly, though they also expected that there would be a tendency for market price to converge towards value over time. We observe that the familiar expectation that a more modern theory is   better than an older one may  not be true in economics, where   conventional neoclassical concepts, theories, and methodologies are still being criticized \cite{keen}.

To physicists, given the dynamics of real economic interactions and the many social and political forces that impinge on economic interactions, the argument that actual trades occur in equilibrium and are market clearing appears strange.  We expect that trades will usually be out-of-equilibrium, that is, price ought to differ from value, and the existence of fluctuating levels of stocks of all kinds of unsold goods is manifest proof that real prices are not ``market clearing". Data show that the so-called {\it ``Law of one price"} is never fulfilled because the market price does not converge towards value even after several centuries \cite{Froot}. Wealth transfer should therefore always occur in real trades because  it is possible to buy or sell a commodity for a price that may be higher or lower than its  value around which actual transaction prices fluctuate.

Moreover, the total amount of wealth that can be transferred in trades from one agent  to the other should be constrained by the total amount of wealth of the poorer trader because a trader cannot (usually) afford to buy or sell a commodity whose value or price  is larger than his/her own total wealth.  
The interesting finding we reach is that because of the above  constraint,  the richer agent always risks less than does the poorer agent in trades. Therefore,  if there is an equal probability to make a ``{\it good} deal" for  both the richer and poorer traders, that is, we assume that all possible outcomes within  the wealth bound are equally likely,  the overall effect is that the entire wealth of a society will eventually concentrate through successive random trades into the hands of the rich; this is a classical statistical problem known as the \textit{gambler's ruin}  \cite{feller}.  This  may explain why it is easier for the rich to get richer and the poor to get poorer over time, than is the reverse.
Of course, in real societies a situation in which the entire wealth concentrates in the hands of  very few people   will cause an economic instability resulting in a social   catastrophe.  A stable society requires that the middle class be the largest.

Indeed, real wealth and income distributions show that the middle class is the largest. The  income distributions, $p\left( w\right) $,     
for very large $w$, almost 1\% of the population (the rich class), satisfy an inverse power
law (IPL) of the form 
\begin{equation}
\lim_{w\rightarrow \infty }p(w)\propto \frac{1}{w^{\mu +1}},  \label{renorm2}
\end{equation}
where, $\mu $ is the IPL index  (note that $\mu $ is  known as  Pareto's exponent \cite{pareto},  of the cumulative distribution $%
P(w)=\int_{w}^{\infty }p(x)dx\propto w^{-\mu }$). Instead, for the low to middle
level of wealth, almost 99\% of the population, the wealth distributions  present a different shape. The distribution curve is non-monotonic and, therefore, there is  a large middle class and a smaller poor class. Wealth or income  distributions at the low-middle range  may
be well fitted, for example, by a Gamma distribution, 
\begin{equation}
\lim_{w\leftarrow \infty }p(w)\propto w^{\eta }\exp [-d~w],  \label{fitt1}
\end{equation}
where $\eta $ and $d$ are appropriate parameters. Figure 1 shows some phenomenological
distributions and cumulative distributions of wealth and income for the United Kingdom and
 the United States \cite{stati}. The 
change of properties between low-middle (gamma-distribution) and high wealth (Pareto's law) ranges is typical of these distribution \cite{souma22,dragulescu,dragulesco2}.

The social economic  catastrophe related to the \textit{gambler's ruin} problem  can be avoided by  assuming a systematic and effectively permanent stochastic bias in trade that favors the poorer trader. This pro-poor bias would move wealth from rich to poor and  reproduce the above socially safer Gamma-like distribution   with a large middle class and a smaller poor class.  We observe that social mechanisms such as the graduated income tax and luxury tax policies are statistically biased in favor of the poor and may play significant roles in moving wealth from  rich to  poor. In addition,  because the main trade that poor to middle wealth individuals make is of their labor services in return for a wage,  we suggest that the wage tends to stochastically exceed the value of labor.   Even if a reader may initially be surprised by our argument that workers' wages should be higher than the value of their labor, we recall that this phenomenon was also observed by  Karl Marx.  In his complete analysis Marx made clear that the wages would normally exceed the value of labor-power \cite{marx1846,marx1857, marx1861} because ``The natural price of labor is nothing but the minimum wage" (p. 55 in Ref. \cite{marx1846}), which is a subsistence payment. Indeed, there should be an economic bias in favor of the workers because workers do not receive only a subsistence payment and, therefore, the wage ``price" should normally exceed wage ``value"! 

In fact, the wage ``price" is not simply the product of the labor itself but  also depends on other factors such as, for example,  the social and political necessity of  stabilizing and favoring the entire society. Moreover, for example, the capitalists would have to  pay a kind of ``social peace" component to avoid strikes because workers might not be  satisfied to receive  only  a subsistence payment. The main means for getting a wage higher than  a subsistence payment is given by the strength of the poor and middle classes that are much larger and, in particular in  modern democracies, politically more influential  than the very small rich class.  The social, political and psychological  effect of the higher economic risk in which the low-middle classes live, due to  their economic resources that are more restricted than those of  the rich, is probably what ultimately  determines  that  real prices and wages are forced to be in an out-of-equilibrium state that favors  the poor-middle classes.

The NSTIM developed herein attempts to simply model the above properties.
The paper has the following structure. In Section II we present some
historical background and indicate what is to be expected from a model based
on complex system theory. In Section III we introduce the nonlinear
stochastic model schematizing the mechanisms of trade and investment,
presumed to underlie the allocation of wealth within a society. In Section
IV we proceed to analyze the NSTIM by computer simulations, under varying
assumptions for the intervening parameters. In Section V the model is shown
to be capable, not only of yielding outcomes which compare favorably with
actual data, but it also sheds some light on the economic basis for
certain overall features of the data. Finally, in Section 6 we present a
detailed discussion of the meaning, in an economic context, of the terms
appearing in the model, and then venture to draw some conclusions on the
implications of the foregoing considerations for the welfare of human
societies.

\section{Some historical observations and asset exchange linear models}

It is straightforward to determine that while income and wealth distributions look qualitatively similar (see Figure 1), they do not coincide because income and wealth are not the same quantity.

The Oxford Dictionary of Economics \cite{oxford} defines \textit{wealth} as
``the total value of a person's net assets, such as, money, shares in
companies, debt instruments, land, buildings, intellectual property such as
patents and copyrights, and valuables such as works of art. From this any
debts owed are subtracted.'' The same source proceeds to indicate that ``the
valuation put on these things is liable to uncertainty and fluctuations, as
many of the assets are not marketed, and those that are may have volatile
market prices.'' The latter aspect will turn out to play a crucial role in
the model. In more quantitative terms a wealth distribution can be obtained
by using the data of all assets and liabilities of a person that must be
reported at his or her death for the purpose of inheritance tax as Dr\u{a}%
gulescu and Yakovenko \cite{dragulescu} did for estimating the wealth
distribution of United Kingdom.
On the other hand, \textit{income} is defined as the amount of money or its
equivalent received during a period of time in exchange for labor or
services, from the sale of goods or property, or as profit from financial
investments. An income distribution is obtained by using data reported each
year to the state for income tax purposes \cite{dragulescu}. Intuitively, it
appears that wealth and income must be somehow related; the variables
wealth, income and, we add, consumption are known, in a society of
interacting agents, to affect each other in a highly non-trivial manner,
through complicated feedback loops, partly described by the economic
concepts of `wealth effect' and `income effect' \cite{oxford,dues}.

Both wealth and income have a long history of being studied as quantitative
indicators of the economic status of a society, at least since the classic
work of Vilfredo Pareto \cite{pareto}, in which he proposed that the
cumulative distribution of the rather sparse income data pertaining to a
number of societies, available to him, could be fitted by an IPL \cite{fn1}. 
With far better data, both in terms of quantity and quality,
being available to later researchers, it is now fairly evident that income
distributions have a more complicated structure, reasonably well described
by  IPLs  only in the uppermost bracket \cite{aoyaha}. There
appears to be a consensus among various scientists \cite
{bouchaud,souma,dragulescu,levy} that the cumulative distribution of wealth,
for the very wealthiest, also exhibits an  IPL  form, while the
data pertaining to other social strata are less certain. An empirical PDF of
wealth increases in the low range, reaches a maximum at middle range and,
finally decreases as an  IPL  at high values of wealth \cite
{dragulescu,stati}. 

Economists have long sought to devise theoretical models that could
reproduce empirical wealth data or, even more ambitiously, that could
`predict' how the distribution of wealth in a society would respond to
changing conditions. None of the existing models are considered entirely
satisfactory \cite{quandrini}. Such efforts have been frustrated by the
previously lamented difficulty in obtaining unequivocal data, especially for
the less wealthy \cite{soto,sands}, but also, we believe, by an approach to
the problem that tends to identify and incorporate a large array of factors
that are believed to affect the dynamics of wealth \cite{quandrini}. Framed
in such reductionistic terms, the problem clearly becomes extremely
complicated. 

An alternative to a direct modeling of all the variables in a
system emerges from a strategy for modeling complex systems \cite{west99}.
The main challenge is to extract the essence of what appears to be a
forbiddingly complicated behavior. In analogy with statistical mechanics, a number of authors have sought to
derive wealth distributions as the steady-state solutions of differential
equations consisting of a stochastic term (investment) and an interaction
term (trade) \cite{bouchaud,souma,solomon,abul,west90}. In these models an
individual's wealth may change in time either, a) because the valuation of
wealth is ``liable to uncertainty and fluctuations'', or b) because wealth
may be exchanged among members of a society. Consequently, we refer to (a)
as \textit{investment} and (b) as \textit{trade}, although such terms can
also be used in a broader sense in these models.
A common trait shared by such models is that the effect of investment is
incorporated through a stochastic multiplicative process term, which is
known to yield solutions with inverse power-law behavior \cite
{bouchaud,west90,sornette}. Differences arise, however, in the interaction terms
describing the effect of trade \cite{bouchaud,ispolatov,hayes}. 

A linear model proposed by Bouchaud and Mezard \cite{bouchaud},  used also by other authors   \cite{souma,solomon,abul} was  borrowed from the
physics of directed polymers and  describes the dynamics of the wealth $W_{i}(t)$
of the agent $i$ in an ideal society of $N$ agents as given by
\begin{equation}
\frac{dW_{i}}{dt}=\eta _{i}(t)~W_{i}+\sum_{j=1(\neq
i)}^{N}J_{ij}~W_{j}-\sum_{j=1(\neq i)}^{N}J_{ji}~W_{i}~. 
\label{firsteq}
\end{equation}
The component $\eta _{i}(t)~W_{i}$ is a Gaussian multiplicative process with variance $\sigma$ that
simulates the investment dynamics.   The two sum terms of Eq. (%
\ref{firsteq}) describe the trade interaction network between the agent $i$
and all other agents in the society and $J_{ij}$ is the linear exchange
rate between agents i and j. This model is solvable in the mean field approximation that implies that $J_{ij}=J/N$ and yields 
\begin{equation}
\frac{dW_{i}}{dt}=\eta _{i}(t)~W_{i}+J(\overline{W} - W_{i})~,
\label{firsteq2}
\end{equation}
where $\overline{W}=N^{-1}\sum_{i}W_{i}$ is the mean wealth and $J$ is the mean field coefficient. The mean-field approximation is useful because Eq. (\ref{firsteq2}) can be associated with a solvable Fokker-Planck equation with the following equilibrium pdf solution
$
p_{eq}(w)=\Psi \exp\left[\frac{1-\mu}{w}
\right]~\frac{1}{w^{1+\mu}}~,
$
where $\Psi =(\mu -1)^{\mu }/\Gamma [\mu ]$ is the normalization
constant and $\mu =1+J/\sigma ^{2}$ is the Pareto exponent
\cite{bouchaud}.

However, despite the elegance of yielding a solvable equation with a solution that succeeds in reproducing a Pareto tail, we believe that  the mean-field approximation obscures any aspect of a realistic trade process. In fact, this approximation implies that in a trade the  agent $i$ gives the $J$ percentage of his own wealth $W_{i}$ to the agent $j$  and in exchange receives the same $J$ percentage of the $j^{th}$ agent's wealth $W_{j}$. This means that in any trade between rich and poor, the poor will always receive an unrealistically large amount of wealth from the rich, and this amount of wealth would  increase if the trader is richer even for the cheapest commodity transferred. Moreover, in the absence of the multiplicative process, the mean-field
approximation  causes the wealth of all economic agents
to exponentially converge toward the mean wealth $\overline{W}$. In fact, the
solution of Eq. (\ref{firsteq2}) without the multiplicative process is
\begin{equation}
W_{i}(t)=\overline{W}+(W_{i}(0)-\overline{W})\exp [-J~t],
\end{equation}
implying that the trade dynamics has the  asymptotic effect of ``equalizing" the wealth
among all members of the society. Therefore, in the absence of investments, there would be neither rich nor   poor! Moreover, even disregarding the mean-field approximation, the linear trade component of  Eq. (\ref{firsteq}) presents the further difficulty of implying that in a trade involving any type of commodity between the same $i^{th}$ and $j^{th}$ agents, the amount of wealth moving from one agent to the other is related to the wealth of both traders and that the linear exchange
rates $J_{ij}$ and $J_{ji}$ are  fixed.  

There also exists a statistical equilibrium theory of markets built on the Maxwell-Boltzmann-Gibbs statistical mechanics and kinetic theory that try to handle the intrinsic stochasticity of the trade phenomenon \cite{dragulesco1,foley,dragulescu,dragulesco2,Chakraborti}.  
 The
kinetic theory approximation  assumes that the transfer of wealth between traders is similar to the interchange  of energy in random elastic collisions between particles and, therefore,   yields  an equilibrium wealth distribution  given by a 
Maxwell-Boltzmann exponential distribution for wealth $p(w)\propto e^{-w/T}$, where $T$ is the ``temperature"  of the market.   According to  such an analogy,  in a trade transaction the total
wealth-energy (or a fraction of it according to the reaction scheme $
[W_{i},W_{j}]\rightarrow [\gamma W_{i}+\varepsilon (1-\gamma
)(W_{i}+W_{j}),\gamma W_{j}+(1-\varepsilon )(1-\gamma )(W_{i}+W_{j})]$, \cite
{Chakraborti}) of both traders should mix and then be randomly split between
the two traders. More recently it was also suggested a kinetic model of market with
random saving propensity that yields to a fixed Pareto index $\mu=1$ \cite{Chatterjee}.

The kinetic theory approach  also presents  problems. In fact, Maxwell-Boltzmann-Gibbs statistical mechanics
 requires the assumption of ``equal likelihood" that necessitates the assumption of ``absence of information" \cite{foley,huang}. These assumptions are problematic   in real trades because  when rich and poor interact, for example,  both agents  are (usually) well aware of their reciprocal social status.  This reciprocal knowledge, in addition to the higher interest that the  poor have in trying to save money or to get more money when they trade with the rich, should constitute the added ``information" that  makes all possible price negotiation outcomes between rich and poor ``not" equally likely. The actual distribution of prices is out-of-equilibrium compared to that based on the equal likelihood assumption of prices \cite{foley} because the information that the traders have about their  reciprocal social status  makes the actual distribution of prices stochastically biased or skewed in favor of the poor regardless of whether the poorer is the seller or the buyer.    
Moreover,  because the kinetic theory assumes that all outputs of the interaction are equally likely, being the conservation of the sum of the two energy-wealths the only constraint,  the poor  would have an extraordinary chance to significantly increase their  wealth when trading with the rich. Finally, the empirical distributions of wealth or income are not monotonic, they increase, reach a maximum and, finally, decrease as an exponential function first and then asymptotically  as an inverse power-law function, see Fig 1. Therefore,  the empirical distribution of wealth or income can not be recovered by the monotonically decreasing exponential distribution of Maxwell and Boltzmann that, at most, is capable of fitting only the middle range of wealth or income distributions \cite{dragulesco2}.

\section{A nonlinear stochastic trade-investment model}

Linear models have unrealistic trade interactions because they do not preserve certain important  properties of  actual trades. 
We  suggest  that Eq. (\ref{firsteq}) be
replaced with a nonlinear model that  preserves the main properties
of a trade mechanism \cite{scafwest}. A trade model has to take into account the role played by 
prices in mediating exchange; how  prices emerge from ideal
negotiations among agents that may belong to different social classes; and finally, how  prices are related to the values of the assets that can be associated only with the wealth of the poorer trader.  In addition, the model should generate a theoretical distribution of wealth to be compared with empirical data
over the entire available range, not just the uppermost bracket where the IPL prevails or the middle range where an exponential behavior seems to prevail \cite{dragulescu}.

 To incorporate these properties of trade 
we
suggest that in the basic trade-investment model  the i$^{th}
$ agent's wealth $W_{i}(t)$ evolves in time according to the discrete nonlinear
stochastic equation 
\begin{equation}
W_{i}(t+1) = W_{i}(t) + r_{i} \xi(t) W_{i}(t)+\sum_{j=1\left( \neq i\right) }^{N}w_{ij}(t).
\label{pd4}
\end{equation}
In this model the investment term is still a multiplicative
stochastic process with Gaussian statistics, as in the Bouchaud-Mezard
model. For convenience, the investment term is here rewritten as $r_{i}\xi
\left( t\right) W_{i}\left( t\right) $ where $r_{i}$, the standard deviation
of the Gaussian variable $r_{i}\xi \left( t\right) $, will be called the 
\textit{individual investment index}. The new element is the quantity $w_{ij}(t)
$, a nonlinear stochastic variable that may change for each transaction and which describes the actual amount of wealth
that is exchanged between the agents $i$ and $j$ in a trade. The
specification of the trade interaction term in the nonlinear stochastic
model requires great care. We observe that our model has only the purpose of  explaining the phenomenon in broad outline not in minute  details; therefore, the interpretation is quite abstract, but we believe realistic.

In a trade
there is a flow of wealth between the two agents only if, in the case
of  barter, the value of the two exchanged assets is different, or, in
the case of  purchase, the value of the asset is different from the price
paid for it. By expressing these concepts in an equation, if the trader $i$
is the seller and the trader $j$ is the buyer, the quantity $w_{ij}$ is
given by 
\begin{equation}
w_{ij}=price - value~,  \label{prival}
\end{equation}
instead, if the trader $i$
is the buyer and the trader $j$ is the seller, the quantity $w_{ij}$ is
\begin{equation}
w_{ij}=value - price~.  \label{prival2}
\end{equation}
If  price and value coincide, the
trade would produce only a transfer of items and money among the agents,
but there would not be any transfer of wealth. If the seller (buyer)
succeeds in selling (buying) an item for a price that is higher (lower) than
the actual value of the item, the seller (buyer) gains wealth from the buyer
(seller). Therefore, we may say that it is the dynamics of making \textit{good or bad deals,}
 in a generalized sense, that continuously generates the random flux of
wealth from one trader to the other in every transaction. 

We observe that the above wealth transfer mechanism  favors one trader and not the other and should not be confused with the fact that  trades occur only if it yields some advantage to both traders.  Without an increase in utility for both traders, the parties would simply refuse to trade. These reciprocal trade advantages are related to the subjective fact that each trader gets what (money or assets) he or she needs and wants more at the occurrence of the trade, while the objective transfer of wealth described by the Eqs. (\ref{prival}) and (\ref{prival2}) is related to the difference between the  price paid for and the value of the asset. Also we observe that if we were interested in the exchange of money instead of wealth, we can proceed with the same above formalism but with $value=0$.

According to Eqs. (\ref{prival},\ref{prival2}) at each update of the economy a random
amount of wealth $w_{ij}$ is positive (negative) if the i$^{th}$ agent is
gaining (losing) wealth in the trade. Consequently, $w_{ij}=-w_{ji}$ and a
typical trade interaction is assumed to follow the scheme 
\begin{equation}
\left[ W_{i},W_{j}\right] \rightarrow \left[ W_{i}+w_{ij},W_{j}+w_{ji}\right]
\label{pd3}
\end{equation}
where the antisymmetric character of the $w_{ij}$ insures that in a trade
transaction the total wealth is conserved. Conversely, wealth may be created
or destroyed  only by investment.

The elements $w_{ij}$ are assumed for simplicity to be Gaussian random
variables with probability density 
\begin{equation}
p\left( w_{ij}\right) =\frac{1}{\sigma _{ij}\sqrt{2\pi }}\exp \left[ -\frac{%
\left( w_{ij}-\overline{w}_{ij}\right) ^{2}}{2\sigma _{ij}^{2}}\right]
\label{pd1}
\end{equation}
where $\overline{w}_{ij}$ is a potential mean wealth that can be transferred between agents $i$
and $j$ and 
\begin{equation}
\sigma _{ij}=hW_{ij}  \label{sdpri}
\end{equation}
is the standard deviation of $w_{ij}$.
The standard deviation of $w_{ij}$ (\ref{sdpri}) is assumed to depend on the
variable 
\begin{equation}
W_{ij}=W_{ji}=\min \left[ W_{i},W_{j}\right]  \label{pd2}
\end{equation}
which implements the assumption that, in a realistic trade, the fluctuation
of wealth involved in a transaction must be  a proper
fraction ($0\leq h\leq 1)$ of the wealth of the poorer trader. In fact, an
individual cannot (usually) buy or sell something  that involves an amount of wealth
larger than his own total wealth. Since the standard deviation of the random
variable, that is, the risk incurred in a trade interaction, is proportional
to \textit{h}, the latter could be interpreted as a \textit{poverty index},
in that in a poorer society a greater fraction of one's wealth would be at
stake in a typical trade. One should also note that the value of $\sigma
_{ij}$ depends on the current status of $W_{i}$ and $W_{j}$ at each update.

The variable $w_{ij}$ of Eqs. (\ref{prival},\ref{prival2}) is stochastic because both the
price and the value of an item may fluctuate. However, we observe
that the value of an item may change only in time because, as explained
above, the \textit{value} is characterized by an  agreement that
involves the entire society and evolves in time according to an investment
mechanism. The price of an item, on the other hand, may change not only in time but it
may be dispersed in space because of its local characterization of being an
outcome of a negotiation that involves only the two actual traders. For
example,  the price of an item may be different at
different locations within the same society at a given time. An explicit
model would have this spatial effect included in several dynamic equations depending on several  variables. Herein, instead,
we imagine that this spatial dependence of the price is described by a unique nonlinear stochastic variable. 
Therefore, the variable $w_{ij}$ of Eq.
(\ref{prival}) is  intrinsically stochastic in nature and can be
associated, for example, with a distribution like Eq. (\ref{pd1}) and not with a
uniquely defined price. These assumptions of the NSTIM are incompatible with the {\it ``Law of
one price''} \cite{lawoneprice}, which is a standard  assumption in  neoclassical economic
models (including Pareto's \cite{mauro}) that attempt to dynamically address
the problem. On the other hand these assumptions are  compatible with the Classical school of thought.

To complete the model we need to specify the value of the mean wealth $%
\overline{w}_{ij}$ to be used in Eq. (\ref{pd1}). A value  $\overline{w}_{ij}=0$ would imply  a statistical equilibrium in which both traders have the same chance to make a good or bad deal. However, as we have explained in the introduction, the mean wealth $\overline{w}_{ij}$ should not be zero because of the many social and political forces influencing economies.  The mean wealth $\overline{w}_{ij}$  is
related to a kind of social differentiation among the members of a society
because the rich and the poor trade in different ways and this keeps the trade in an out-of-equilibrium state. We  assume that
the poor are constrained by their poverty to be more careful in their trades
and therefore, for example, they may often look for the best price
opportunity for saving money. On the other hand, because of their economic
strength, the rich may be willing to purchase items regardless of cost. This
is a realistic assumption in particular when the value of an item is low
compared with the total wealth of the trader, that is, for example when a
wealthy agent would like to buy something from a poorer agent. Of course, if
the wealth involved in a trade is compatible with the agent's wealth, he or
she is expected to be more prudent. In other words, a trade can take place only if the two agents, through a
negotiation, reach an agreement about the price of the asset. The trade
transaction has a higher probability to occur if the price is below a
threshold at which the buyer is willing to buy. Because this threshold
increases with the total wealth of an agent, when a wealthy agent would like
to buy something from a poorer agent there is a higher probability that the
transaction occurs at a higher price than when a wealthy agent would like to
sell the same item to a poorer agent. This asymmetric disadvantage-advantage 
tends to disappear when the two traders are economically equivalently. 

In conclusion, in the NSTIM the volume of wealth
involved in a trade between two agents is related to the wealth of the
poorer of the two traders, so we assume that $\overline{w}_{ij}$ is related
to the value $\sigma _{ij}.$ In this way $\overline{w}_{ij}$ is of the same
magnitude as the amount of wealth involved in the trade, and it is related
to the total wealth of the two traders through the nonlinear expression 
\begin{equation}
\overline{w}_{ij}=\alpha _{ij}~h~W_{ij}.  \label{pd5}
\end{equation}
The variable $\alpha _{ij}$ is given, for example, by the nonlinear term that measures the out-of-equilibrium status of the trade, 
\begin{equation}
\alpha _{ij}=f\frac{W_{j}-W_{i}}{W_{i}+W_{j}}.  \label{pd6}
\end{equation}
The quantity $f>0$ is called the \textit{social equality index}, and measure the strength of  the
bias in favor of the poorer trader. In fact, according to Eqs. (\ref{pd1})
and (\ref{pd6}), if the wealths $W_{i}$ and $W_{j}$ of the two traders are
almost the same, then $\alpha _{ij}\approx 0$ and both traders have an equal
chance of gaining or losing wealth. If, instead, for example, $W_{j}\gg
W_{i} $, we have $\alpha _{ij}\approx f$, the distribution $p(w_{ij})$ (\ref
{pd1}) is out-of-equilibrium, that is, shifted toward positive values and the trader $i$, the poorer
trader, has a better chance to gain wealth in the trade. We stress that with this mechanism, in trading with the rich the poor can have only a realistic ``better chance" to improve their own wealth by an amount related to their own restricted resources,  and not, as the mean field and kinetic theory  approximations would  imply,   have the possibility of gaining an unreasonably large amount of wealth from the rich.

\section{Numerical simulations}

We study the properties of NSTIM numerically and initiate all simulations by assuming an ideal society consisting of 10$%
^{5}$ agents, with wealth uniformly distributed among them. 

The first case we examine is with  $h>0$, $f=0$ and $r=0$, that is, we assume
a symmetric trade-alone economy where $\overline{w}_{ij}=0$ in ($\ref{pd1}$)
for all $j$ and $i$. Such a choice implies that  in any trade each amount of transferred wealth is
equally likely to be gained or lost by any given agent. The simulation shows that
under such conditions, the available wealth is rapidly concentrated in the
hands of a relatively few agents. Moreover the inequality in the
distribution of wealth increases with the number of iterations, as shown in
Figure 2. Such an outcome is not surprising; in fact, if one interprets the
standard deviation $\sigma _{ij}$ as the \textit{risk} incurred in a trade
by agents $i$ and $j$, it is clear that the poorer agent is taking a greater
risk in proportion to his/her wealth, with the disparity increasing the
greater the difference in wealth between the traders.
The dynamics of a system such as that represented by the symmetric trade-alone
model are unstable, in the sense that relative differences generally tend to
be further amplified. If we set the threshold of `economic death' at some
finite level $W_{j}>minimum$, implying that an agent with such little wealth
cannot continue to trade, the situation becomes completely analogous to the
classical problem of the \textit{gambler's ruin} \cite{feller}. Modern societies generally avert such
catastrophes \cite{fn2}, thereby implying the unrealistic nature of trades in a statistical equilibrium between rich and poor
and  suggesting the need for a mechanism which would dampen
differences between rich and poor, rather than amplify them. 

The second case we examine is with  $h>0$, $f>0$ and $r=0$,
The quantity $f>0$ measures the
bias in favor of the poorer trader. This version of the trade-investment
model clearly implies that agent \textit{i} has a greater chance to draw a
positive value for $w_{ij}$, representing a gain, if agent $i$ is poorer
than agent $j$, and this chance increases with increasing economic disparity.
 In Figure 3 we show the resulting
probability density functions for different values of the social equality
index $f$ for a given value
of $h$. The distribution of wealth becomes narrower with increasing $f$,
meaning that the probability of being rich falls off rapidly with increasing
wealth. Therefore, the condensation of wealth is less pronunced in a society characterized by a high value of the social equality index $f$ that would imply a higher chance for the poor to gain wealth from the rich. Instead,  Figure 4 shows that for a given value
of $f$, increasing the poverty index $h$ leads to greater economic
disparity and, therfore, to a higher condensation of wealth. The interpretation is that in a poorer society, and one based on
trade alone, the potentially ruinous effects of an unfavorable trading
sequence are more pronounced, and this will benefit the rich against the poor.
All simulations shown in Figures 3 and 4 are well-fit by the Gamma-like
distribution 
\begin{equation}
p\left( w\right) =a\left( w-c\right) ^{\eta }\exp \left[ -dw\right]
\label{dp7}
\end{equation}
which clearly exhibits a non-monotonic behavior with an exponential-like
tail.  In Table 1 we display the values of the parameters utilized to fit the
outcomes of different simulations.
To summarize, the latter version of the asymmetric or out-of-equilibrium trade-alone model ( 
$h>0$ and $f>0$, $r=0$) appears more realistic than the previous one with $f=0$ in
two respects: 1) it yields, at least qualitatively, the emergence of a
sizeable `middle-class' as opposed to an overwhelmingly rich elite and a
vast population of paupers as did the previous model; 2) it admits
well-behaved stationary solutions, whereas $f=0$ would lead to an inexorable
drift towards a singularity; 3) the trade mechanisms gives origin to a Gamma distribution that is what is observed for the low-middle classes in phenomenological data, as Figure 1 shows. In fact, the economy of the low-middle classes is presumably characterized mostly by trades.

The third case we examine is with  $h>0$, $f>0$ and $r>0$,
which means we now examine the role played by the multiplicative processes
that describes investments in the model economy. In Figure 5 we show the
outcome of simulations with both mechanisms, investment and trade,
simultaneously active. The curves highlight the effect of investment by
plotting three probability density curves, one for each value of $r_{j}$,
for fixed values of $f$ and $h$. It appears that increasing $r$, all other
things being equal, leads to a broadening of the wealth distributions and
thus, to a greater economic disparity. The economic interpretation is that
greater disparities arise in a society wherein one's worth tends to
fluctuate more, either because its members deliberately place more of their
wealth in riskier assets (investments with greater potential payoff engender
more risk) or because the valuation of their assets is intrinsically more
erratic due, for example, to uncertain times. As in the previous case, if we keep $r$ fixed, the model simulations show that the Pareto index increases with increasing $f$ and decreases with increasing $h$. 

The investment process is implemented in the simulation by imposing that
every 10$^{4}$ trades the wealth of all agents is reinitialized by the
multiplicative process. In Figure 5 is visible a partial fit of the tail of
one of the computer-generated probability density functions with a Pareto
distribution $x^{-\delta }$, where $\mu =\delta -1$ is the Pareto exponent.
It is evident from the figure that, as expected, the inclusion of the
multiplicative process term is capable of generating distributions endowed
with inverse power-law tails. We find that the computer-generated
probability density functions may apparently be well-fitted over the entire
range by functions of the form 
\begin{equation}
p\left( w\right) =\frac{a\delta w^{\gamma }}{\left( 1+bw\right) ^{\delta
+\gamma }}  \label{pd8}
\end{equation}
Table II records the fitting parameters for the two wealth distributions
denoted by triangles and stars in Figure 5.
The cumulative probability function, that is, the integral of the
probability density function depicted in Figure 5 by circles, is shown in
Figure 6, along with an IPL fit for the upper wealth region.
In the inset is shown the fit of the middle range of the same curve
implemented with an exponential function proving that the low-middle range of the distribution can be also well fit with a Gamma distribution.
In fact, in some simulations the distribution separation between the low-middle range and the upper range emerge more clearly. 

At this stage, one can conclude that the NTSTI model (\ref{pd4}),
implemented with both investment and trade mechanisms, and with the latter
biased in favor of the less wealthy agent, can successfully reproduce
important qualitative features observed in the real world, namely, the
stratification of society into a poor class, a large middle class and an
affluent but very small upper class. A trade-alone mechanism would be sufficient to get
such a stratification, it  leads to a Gamma distribution as observed in phenomenological data regarding  the low and middle classes, see Figure 1b. To
obtain the observed Pareto's law, that is, an IPL tail for the
uppermost  class it is necessary to include the investment mechanism.

\section{Comparison of calculations with data}

In this section we  assess the ability of the NSTIM to
reproduce quantitative details of empirical data, see Figure 1, as opposed to merely
reproducing general qualitative features of data. In a recent empirical
study, Dr\u{a}gulescu and Yakovenko \cite{dragulescu} conclude that the data
pertaining to the cumulative distribution of wealth in the United Kingdom
and income for the United States can be well fitted by an exponential
(Maxwell-Boltzmann) function in the low-middle range, followed by an inverse
power-law (Pareto) tail in the high-end region, as we also obtained from the
preceding simulations. However, as these authors show, such an exponential
behavior cannot be correct over the entire domain of the data because the
PDF of wealth is not monotonic. Dr\u{a}gulescu and Yakovenko recognize that
in the low-middle range the wealth PDF has been well fitted by a Gamma
distribution, as the trade-investment model produces. In any case, it turns
out that a cumulative distribution of wealth has been obtained by appeal to a
variety of mechanisms. For example, Huang and Solomon \cite{huang2} ascribe
the appearance of such dual behavior in wealth distributions to the fact
that an idealized marketplace has a finite size; Souma et al.\cite{souma}
identify the source of these features with the rewiring of links in a
small-world network \cite{watts} over which a stochastic process is assumed
to evolve and finally Montroll and Shlesinger \cite{montroll2}, in their
discussion of income, arrive at this result using renormalization group
scaling arguments and these arguments can be simply recast in terms of
wealth.

To understand why the distribution of wealth for the rich is different from
that of the rest of society, we observe that they rely on different economic
instruments: the rich mainly on investments and all the others mainly on trades. This plausible statement has the virtue of empirical evidence 
\cite{baekgaard}. A simple way to incorporate this dichotomy into
the NSTIM is to subdivide the \textit{N} agents into two groups with
different investment indices, 
\begin{eqnarray}
\Delta W_{i} &=&r_{1}\xi W_{i}+\sum_{j=1\left( \neq i\right) }^{N}w_{ij}%
\text{ ; }1\leq i\leq N_{1}  \label{pd9} \\
\Delta W_{i} &=&r_{2}\xi W_{i}+\sum_{j=1\left( \neq i\right) }^{N}w_{ij}%
\text{ ; }N_{1}\leq i\leq N  \label{pd10}
\end{eqnarray}
where $N=N_{1}+N_{2}$ and $r_{1}\neq r_{2}.$ Figure 7 shows two cumulative
distributions obtained with the parameters: social equality index $f=0.3$,
poverty index $h=0.05$ and investment indices for 50\% of the population $%
r_{1}=0.075$ and 0.055, while $r=0$ for the other half in both simulations.
It is clearly visible in the figure that this artificial two-tiered
subdivision of society accentuates the distinction between an IPL tail
region (about 1\% of the population) of the wealth distribution and another
region (about 99\% of the population) with completely different curvature that can be fitted with a cumulative gamma distribution. 
Figure 7 shows  that by choosing opportune parameters of the model it  is possible to fit phenomenological  distributions quite well.
In summary, the distributions of wealth and income show an anomalous shape well described by a Gamma distribution at low and middle wealth and a Pareto's tail at high wealth, that is, with a complex shape  that may be approximately described by Eqs. (\ref{renorm2}) and (\ref{fitt1}).

We interpret this result as a
signal that such a duality of economic mechanisms, pursued by different
strata of society, may indeed be responsible for the observed dual behavior
of empirical curves. The Pareto exponents of the two fitting curves in Figure
7a are, respectively, $\alpha =1.5\pm 0.02$ and $2.5\pm 0.02$, which straddle
the empirical values as those shown in figure 1a ($\alpha =1.85$) about UK \cite{dragulescu} and those shown in figure 7b about the cumulative distributions in USA during 1980 ($\alpha =2.2$) and 1989 ($\alpha =1.63$). Furthermore, increasing 
$r$ for the investment-prone part of society leads to smaller Pareto
exponents, as expected, and thus to greater economic disparity.

\section{Discussion and Conclusions}

Naturally, it would be an arduous task to measure the values of the
fundamental parameters $r$, $h$ and $f$ of the NSTIM model in real economic
situations. The actual values of these parameters are expected to fluctuate
around average values and they may assume arbitrarily complicated
functional forms to account for varying structures of society's trade and
investment networks. Nevertheless, from a complex systems viewpoint, a framework has
been constructed wherein general features observed in empirical wealth
distributions can be ascribed to specific activities that are known to be
present in society and, perhaps more importantly, tests can be run on the
social impact of variations, which are known to
occur, of specific economic factors.

Phenomenological distributions of wealth show an abrupt change
between the low-middle classes (99\% of the population), where a Gamma distribution dominates, and the uppermost class (1\% of the population) where a Pareto's distribution dominates. We have shown that the NSTIM model
consistently  interprets such  behavior as due to the fact that trade is the
main mechanism that characterizes the low and middle classes and this produces a Gamma distribution of wealth, whereas
investment,   is the main mechanism that influences the uppermost class individuals  who own the means of large production. The trade mechanism depends on two parameters: a poverty index $h$, and a social equality index $f$.
We saw that by increasing $h$ or decreasing $f$ wealth condenses more easily. The investment process  can be modelled with a multiplicative stochastic mechanism whose strength is measured by the parameter $r$ and  generates a Pareto's law. By incresing the parameter $r$ wealth condenses more easily because  the economy of
the low-middle classes remains mainly characterized by trades even though some
investment can be present within these classes. Of course, the three parameters may change in time and this explains why the Pareto index is not constant but evolves \cite{souma22} as Fig. 7b shows.   

As a simple interpretative application of the model, herein  we attempt an explanation of a sociological phenomenon that is easily observed in history and that was summarized by Pareto in the theory known as ``{\it the circulation of the elites}" \cite{pqretoelit} and in his criticism of the Marxist revolutionary theory \cite{paretomarx}. In fact, in poor societies where modern means of production are for
whatever reason absent, or simply insufficiently capable of producing
surplus value, revolutionaries usually think that alternate means should be developed to
redistribute wealth by force or by deceit with the results of increasing the social equality index  $f$. But the success of such policies would be doubtful. In fact, a
society without sufficient means of production would be characterized by a
high poverty index $h$ that  statistically favors a small rich class in
their trade with the poor. On the other hand  the social equality index  $f$ can not be  arbitrarily increased because  too high an $f$ would excessively reduce the capital that is necessary for  production and would cause a further impoverishment of  the society. 
Moreover, the ensuing insecurity of a revolution
would be reflected by a concomitant increase of the volatility parameter $r$, which would again favor
a small upper class.
 Therefore, the NSTIM model seems to confirm the realistic expectation that without  consistent economic development, which reduces the poverty index $h$ that contrasts the wealth condensation, a revolution would simply have, at most, the asymptotic effect of  supplanting
one ruling elite with another because there would not be a significant 
redistribution of wealth, despite Marxist promises to the contrary \cite{paretomarx}. 

Finally, we owe the reader a wider discussion about  the wisdom of opting for a model of
trade that favors the poor against the rich, even if such a bias is only statistical. Without such a bias the rich would get richer and the poor would get poorer over time.
The mechanism that in a trade favors the poorer of the two agents  is possible because in a real economy, contrary to an assumption of the Neoclassical school of economics,  price and value of a commodity do not coincide: this is   what generates the transfer of wealth from one trader to another according to Eqs. (\ref{prival}) and (\ref{prival2}). The price dispersion is contingent on the trade phenomenon itself and not simply a  transition to an equilibrium price as the {\it ``Law of one price"} would suggest \cite{lawoneprice}. In fact, the price dispersion   is produced by different possible  outcomes of  economic  negotiations between pairs of traders and it is associated with several factors, among them the most important being the economic  differentiation among the social classes yielding a  price discrimination mechanism \cite{pricedisc} and, therefore  a divergence between price and value, the latter defined as the equilibrium price. 

The social differentiation  induces asymmetric behaviors that yield  statistically opposite outcomes  when two agents, belonging to different social classes, negotiate a transaction; for example, the poor try to sell their services and products at the highest prices and purchase at the lowest prices. By including such a mechanism, the nonlinear stochastic trade model  takes into account the importance of the role of prices in mediating exchange and redistributing wealth among members of the society. NSTIM also  takes into account how the price emerges from a negotiation mechanism whose stochastic outcome can be only a fraction of the wealth of the poorer party and can not be related to the richer trader's wealth. This is because a trader cannot (usually) afford to buy or sell a commodity whose value is larger than his/her own total wealth.  Instead, the probability of making a ``good deal" depends nonlinearly on the wealth of the two traders and is statistically biased in favor of the poorer agent. In fact,  the richer party is  less risk averse when bargaining over a given amount, than is the poorer party. This means that the poorer party should be a stronger bargainer than the rich to get the better of the deal. 

In particular,  in real societies the main trade transactions are those taking place between citizens and
the state through the tax system, and between employers and employees.  It is easy to realize that these processes  are statistically biased in
favor of the poor and play a significant role in redistributing wealth.  
The graduated income tax  requires that the rich 
 pay a higher percentage of taxes than do the poor. There are also luxury taxes for products that only the rich can afford to buy. On the contrary, there are several tax reductions for products necessary to everybody such as, for example,  food staples.
 These mechanisms are
devices deliberately put in place to ameliorate the economic gap between
rich and poor.  In fact, the tax system can be interpreted as a trade
through which the citizens buy services from the state. Because the services
granted by the state are expected to be proportional to a citizen's wealth,
the graduated tax system has the effect of discriminating among the
citizens, forcing the wealthy class to pay a higher price for services than
that paid by the less wealthy.
About the relation between employers and employees, we notice that  every time Marx spoke of the
relationship  between  the  wage  and the value of labor-power, he used the
term  ``minimum  wage" \cite{marx1846,marx1857, marx1861}, thus emphasizing that in practice he expected the wage
to  exceed  this  minimum and hence there to be a price-value divergence in
favor  of  the  working  class at the expense of capitalists. These effects can be incorporated into the social equality
index $f$ of Eq. (\ref{pd6}) that measures the statistical bias of the trade
in favor of the poor.

We observe that according to the perspective of the Neoclassical school of economics, the out-of-equilibrium concept that wage is different from the value of labor makes no sense. The Neoclassical model of the labor market presumes that the wage equals the marginal product of labor (see Keen 2002 Ch. 5 \cite{keen}), and wage value and wage price therefore coincide. However, from a Classical point of view, the concept is quite different. The Classical school defined the value of any commodity as its cost of production, and the cost of production of the commodity labor is therefore the means of subsistence. Our finding that trade is stochastically biased in favor of the poor is consistent with the fact that the wage price can be systematically higher than the means of subsistence and, therefore, the workers receive more than a subsistence wage. 
This empirical conclusion is not inconsistent with a Neoclassical perspective on the relationship between the actual wage and the means of subsistence, since in the Neoclassical theory there is no relationship between the marginal product of labor and a subsistence level of income. But the conclusion that there is a transfer of wealth from rich (employers) to poor (workers) via  worker-biased wages is inconsistent with the vision of the labor market   (and, therefore,  all other markets)  being in equilibrium.

In conclusion, we believe that the most important theoretical aspect of our findings is that our study allows us to  compare two rival economic theories (Classical against Neoclassical) of price and wealth distribution, and seems to come down on the side of the Classical School concerning the importance of the role played by  price/value divergence. This divergence  is alien to
Neoclassical  economics,  which  has  melded  price  and value into  one
thing.  Neoclassical  analysis  is  predicated  upon  equilibrium,  and  in
equilibrium  price  equals  value, and no transfer of wealth is possible in trades.  
The  preceding  Classical  School  did  have a theory in which market price
could  and  normally  did  differ  from value as seems phenomenologically evident. The Classical School defined
value  effectively  as the cost of production of a commodity; the price was
what  it  sold  for on the market. They expected price/value divergences to
occur  all  the  time and that, therefore, out-of-equilibrium  trades were possible. Our findings stress the importance of  out-of-equilibrium  mechanisms that bias the trade transactions  in favor of the poor because  they are not only possible but they seem necessary to stabilize society by avoiding the economic catastrophe whereby the entire wealth  concentrates in the hands of very few rich people.

{\ {\large \textbf{Acknowledgment:}}}\newline
N.S. thanks the Army Research Office for support under grant DAAG5598D0002. We would like to thank prof. M. Boianovsky, prof. T. Groves and prof. D. K. Foley  for some useful discussions. We would like to thank prof. S. Keen for some very useful suggestions.

\newpage
\begin{figure*}
\epsfig{file=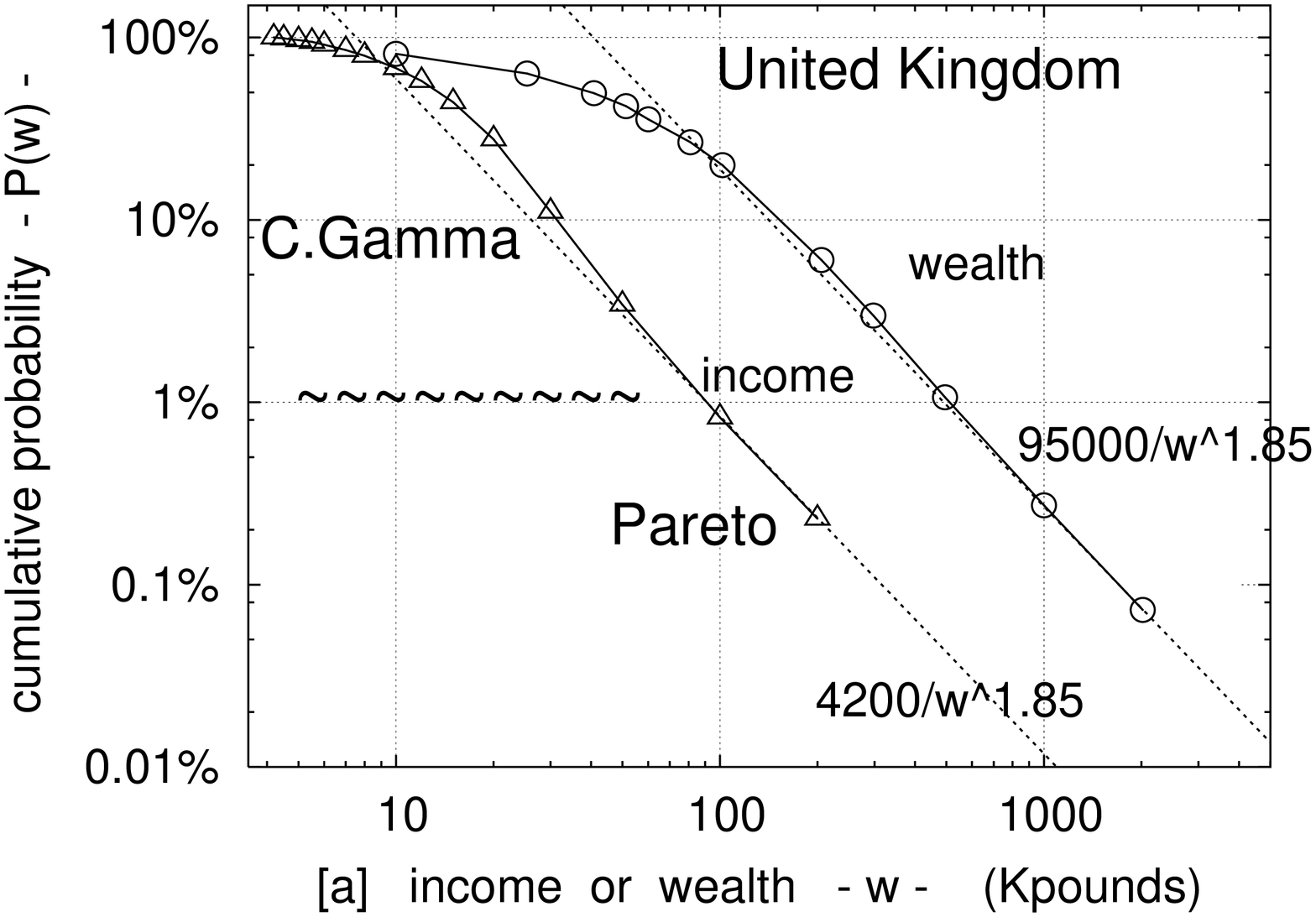,height=8cm,width=6cm,angle=-90}
\epsfig{file=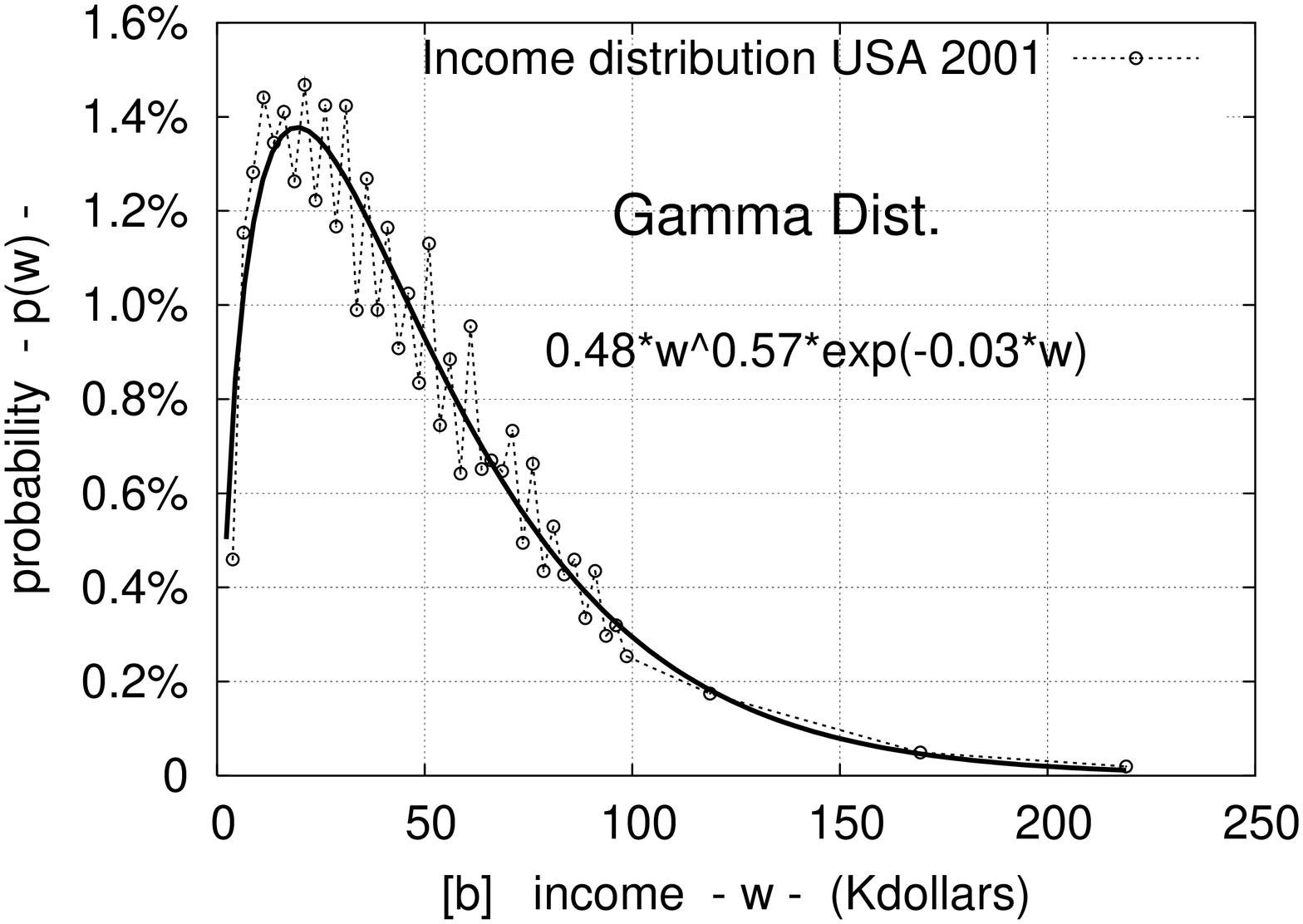,height=8cm,width=6cm,angle=-90}
\caption{[a] Cumulative wealth (1996) and income (1998-1999) distributions in United Kingdom;  the upper class (almost 1\% of the population) the distribution is well fit with a Pareto's law. [b] Income distribution (2001) in USA; the low-middle classes (almost 99\% of the population), the distribution is well fit with a Gamma distribution, Eq. (\ref{dp7}).  }
\end{figure*}

\newpage 
\begin{figure*}[tbp]
\epsfig{file=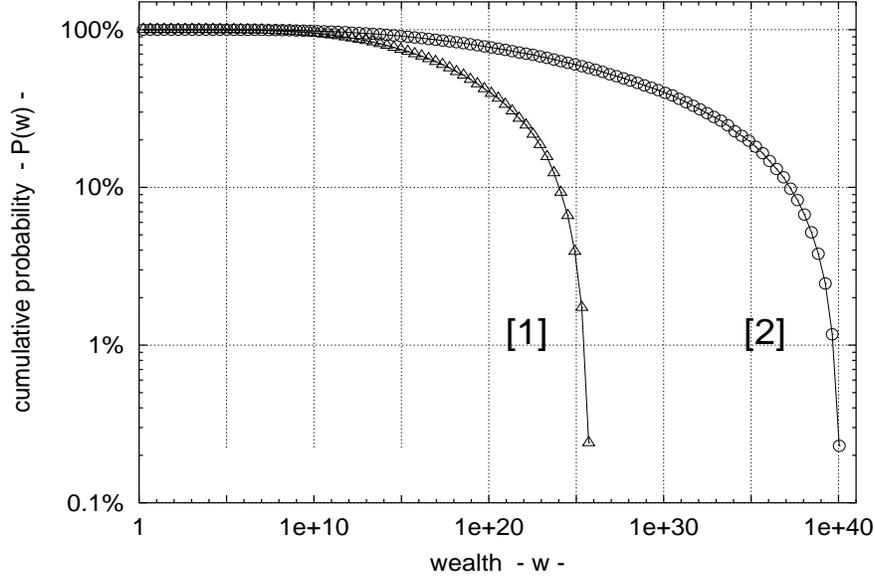,height=12cm,width=8cm,angle=-90}
\caption{Cumulative wealth distribution for the symmetric trade-alone model.
The indexes are: $h=0.05$, $f=0$ and $r=0$. Case [1] is after 100 million
trade-interactions, and case [2] is after 200 million trade-interactions.
The initial wealth distribution is uniform. The figure shows that this model
yields to a huge wealth gap between the rich and poor that increases with
the number of interactions. The wealth is measured in units of the poorest
agent's wealth. }
\end{figure*}

\newpage 
\begin{figure*}[tbp]
\epsfig{file=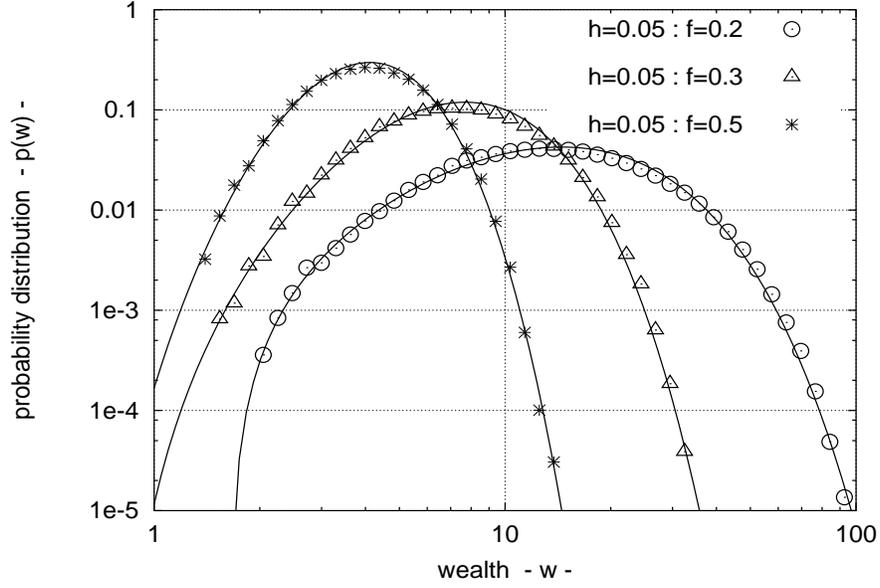,height=12cm,width=8cm,angle=-90}
\caption{Wealth probability density for the asymmetric trade-alone model
with a fixed poverty index $h=0.05$. The wealth condensation increases by
decreasing the social index $f$. The investment index is $r=0$. The
distributions are fitted by a Gamma distribution Eq. (\ref{dp7}). The
fitting parameters are in Table I. The wealth is in units of the poorest
agent's wealth. }
\end{figure*}

\begin{table}[tbp]
\begin{tabular}{|c|c|c|c|c|}
\hline
& a & d & $c$ & $\eta$ \\ \hline
f=0.2 & 3e-3 $\pm$ 3e-4 & 0.147 $\pm$ 2e-3 & 1.64 $\pm$ 0.05 & 1.9 $\pm$ 0.05
\\ \hline
f=0.3 & 5e-3 $\pm$ 1e-3 & 0.56 $\pm$ 0.02 & 0.8 $\pm$ 0.1 & 3.9 $\pm$ 0.2 \\ 
\hline
f=0.5 & 0.10 $\pm$ 0.06 & 1.9 $\pm$ 0.06 & 0.5 $\pm$ 0.1 & 6.8 $\pm$ 0.5 \\ 
\hline
\end{tabular}
\caption{Fitting parameters of Eq. (\ref{dp7}) for the trade-alone economy
(r=0), see Fig. 3. The poverty index is fixed $h=0.05$.}
\end{table}

\newpage 
\begin{figure*}[tbp]
\epsfig{file=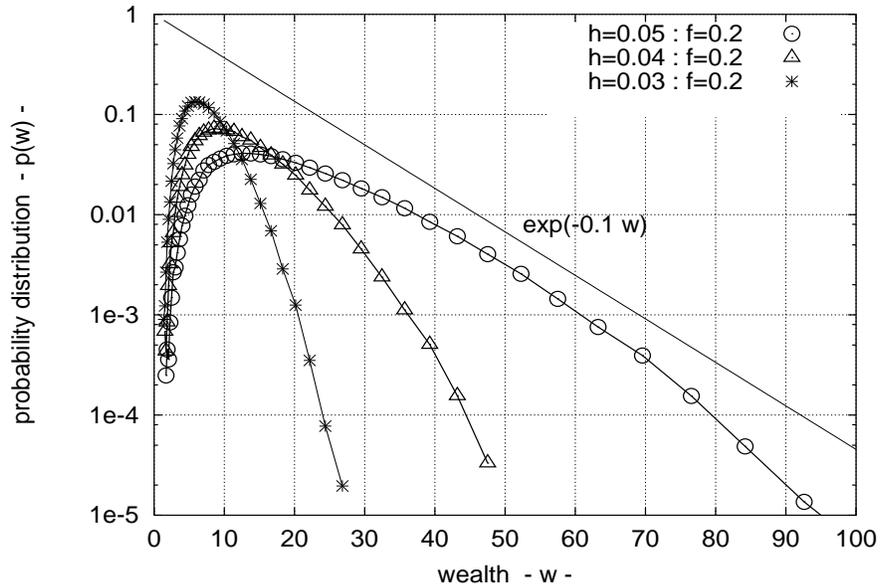,height=12cm,width=8cm,angle=-90}
\caption{Wealth probability density for the asymmetric trade-alone model
with a fixed social index $f=0.2$. The wealth condensation increases by
increasing the poverty index $h$. The distributions are compared to an
exponential Maxwell-Boltzmann distribution. The wealth is in units of the
poorest agent's wealth. }
\end{figure*}

\newpage 
\begin{figure*}[tbp]
\epsfig{file=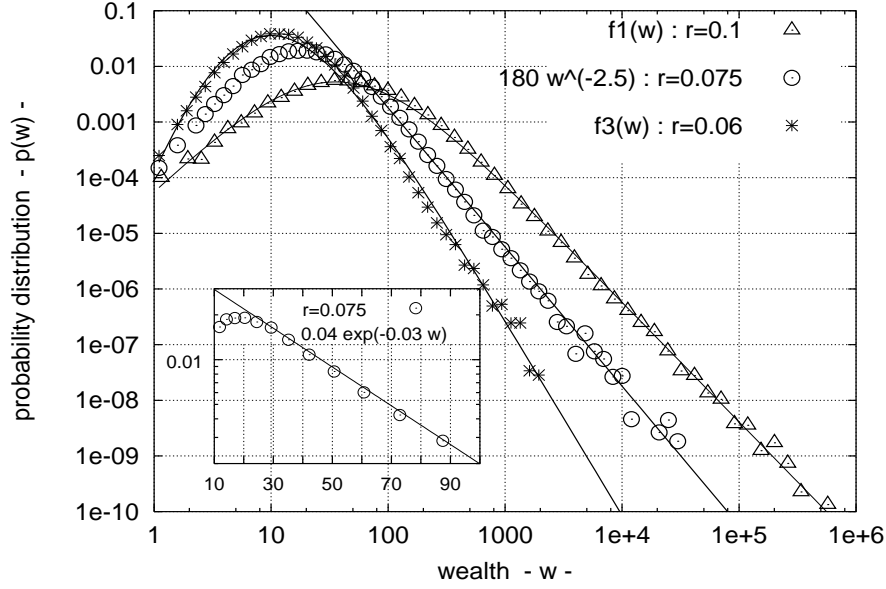,height=12cm,width=8cm,angle=-90}
\caption{Trade-investment economy. The social and poverty index are fixed; $%
f=0.3$ and $h=0.05$. The probability distributions (tringles) and (stars)
are fitted by using Eq. (\ref{pd8}). The fitting parameters are in Table II.
The tail of the probability distribution (circles) is fitted by a power law
of the type $1/x^{\mu+1}$ where $\mu=1.5$ in the Pareto's exponent. The
small picture shows the probability distribution (circles) in the interval
[10:100] that can be apparently fitted with an exponential Maxwell-Boltzmann
distribution but not at low values of $w$. The wealth is measured in units of the poorest agent's wealth. 
}
\end{figure*}

\begin{table}[tbp]
\begin{tabular}{|c|c|c|c|c|}
\hline
& a & b & $\gamma$ & $\delta$ \\ \hline
f1(w) & 6e-5 $\pm$ 1e-5 & 2.6e-2 $\pm$ 2e-3 & 2 $\pm$ 0.1 & 2.15 $\pm$ 0.05
\\ \hline
f3(w) & 5e-4 $\pm$ 1e-4 & 0.17 $\pm$ 0.02 & 5.8 $\pm$ 0.6 & 3.5 $\pm$ 0.1 \\ 
\hline
\end{tabular}
\caption{Fitting parameters of Eq. (\ref{pd8}) for the Trade-investment
economy, see Fig. 5. The social and poverty index are fixed; $f=0.3$ and $%
h=0.05$.}
\end{table}

\newpage 
\begin{figure*}[tbp]
\epsfig{file=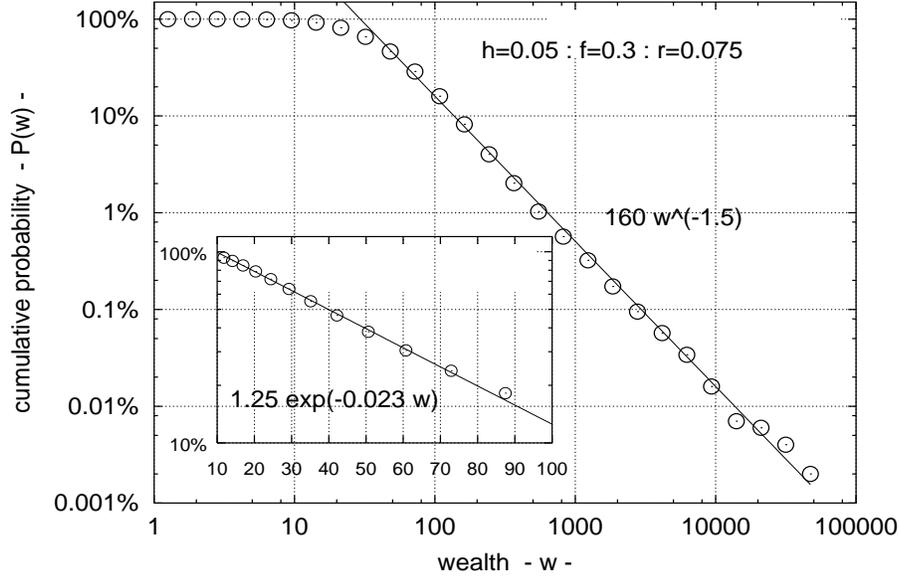,height=12cm,width=8cm,angle=-90}
\caption{Cumulative probability for an trade-investment economy with $h=0.05$%
, $f=0.3$ and $r=0.075$. The Pareto's exponent is $\mu=1.5\pm0.02$. The
little picture shows that in the interval [10:100] the P(w) can be
apparently fitted by an exponential Maxwell-Boltzmann distribution. The
wealth is measured in units of the poorest agent's wealth.}
\end{figure*}

\newpage 
\begin{figure*}[tbp]
\epsfig{file=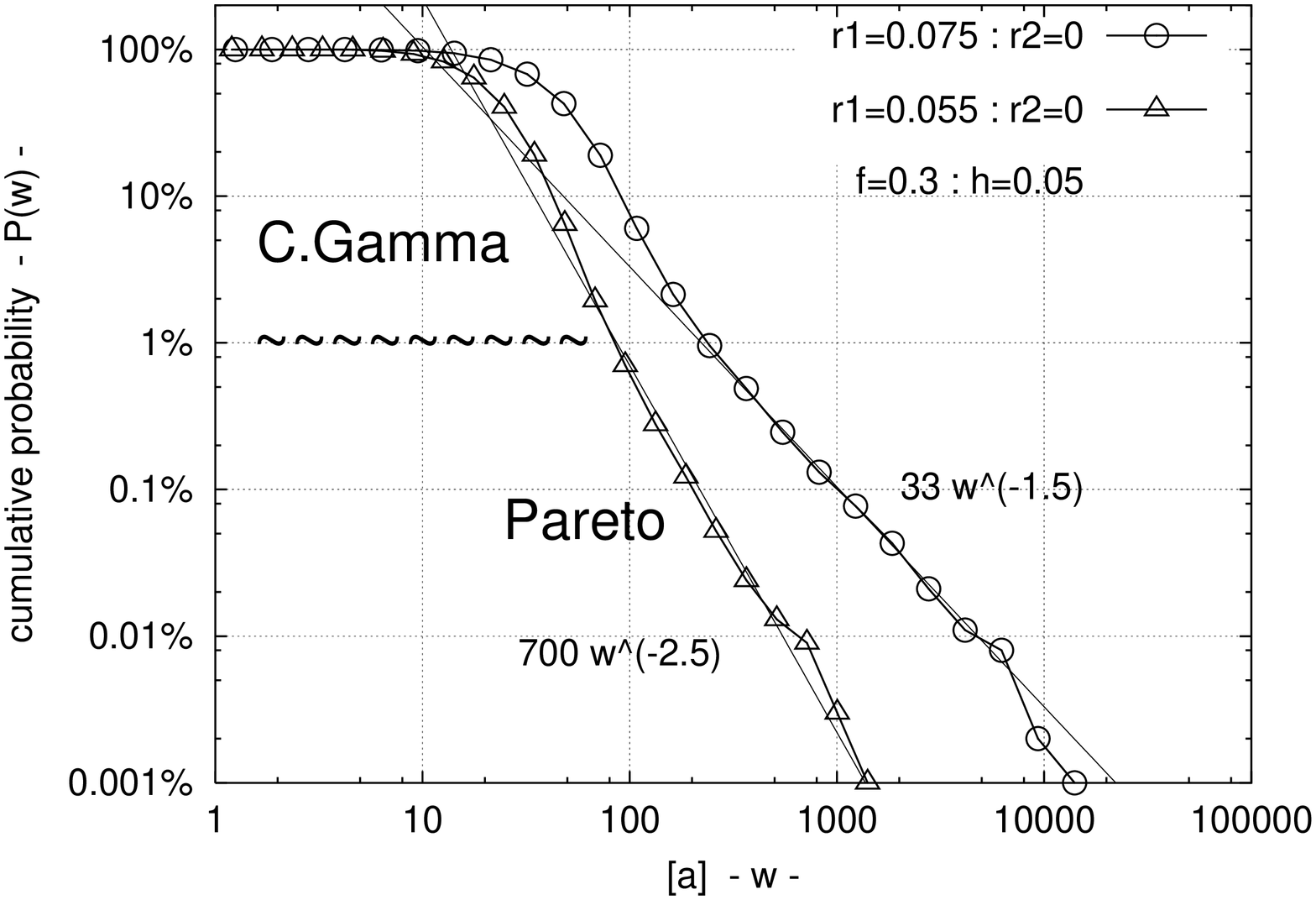,height=8cm,width=6cm,angle=-90}
\epsfig{file=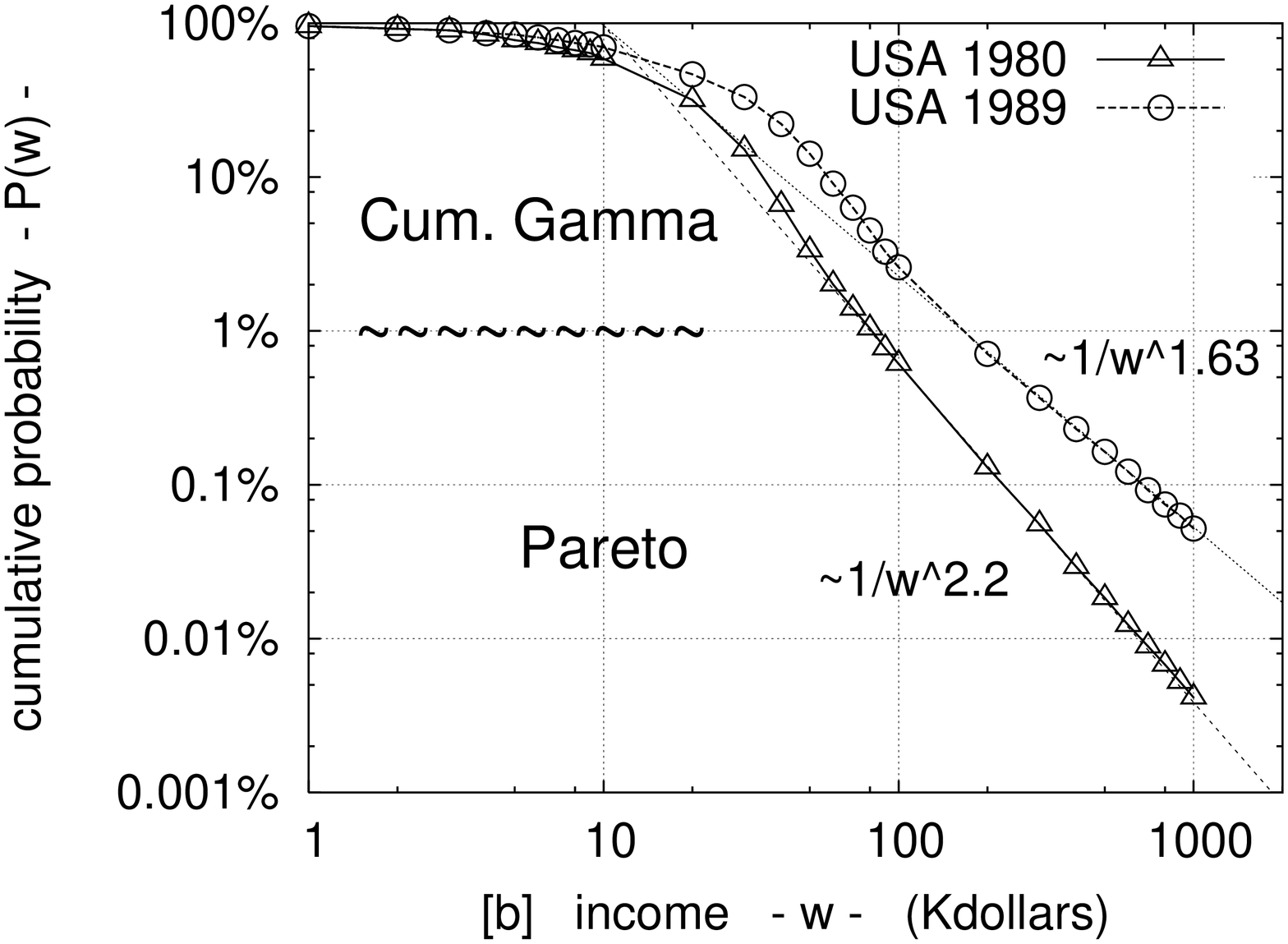,height=8cm,width=6cm,angle=-90}
\caption{[a] Cumulative probability for a double trade-investment economy. One
half of the population has the investment index $r_2=0$, the other half of
the population has in one case $r_1=0.075$ and in the other $r_1=0.055$. In
both cases the social index is $f=0.3$ and the poverty index is $h=0.05$.
The Pareto's exponents are $\mu=1.5\pm0.02$ and $\mu=2.5\pm0.02$. While 99\% of the population follow a cumulative Gamma distribution and 1\% of the population follow a Pareto law. The wealth
is measured in units of the poorest agent's wealth. [b]  Cumulative income distributions in USA during 1980 and 1989.}
\end{figure*}


\begin{thebibliography}{99}


\bibitem{bouchaud}  J.P. Bouchaud and M. Mezard, Physica A {\bf 282}, 536 (2000).

\bibitem{dragulesco1}  A. Dr\u{a}gulescu and V. M. Yakovenko, Eur. Phys. J. 
\textbf{B17}, 723-729, (2000).

\bibitem{foley} D. K. Foley, J. Econ. Theory {\bf 62}, 321 (1994).

\bibitem{lawoneprice}  An economic rule which states that in an efficient
market, a security must have a single price, no matter how that security is
created. For example, if an option can be created using two different sets
of underlying securities, then the total price for each would be the same or
else an arbitrage opportunity would exist. 
About the deviation from the {\it ``Law" of one price}, see also Ref. \cite{Froot}. 

\bibitem{say} S. Kates, {\it Two Hundred Years of Say's Law: Essays on Economic Theory's Most Controversial Principle},
Edward Elgar Pub, (2003). For an original critique of the ``Say's Law" see chap. 12, by S. Keen.  


\bibitem{keen} S. Keen, {\it Debunking Economics: The Naked Emperor of the Social Sciences}, Zed Books (2002). 

\bibitem{Froot} K.A. Froot, M. Kim and K. Rogoff ``The Law of One Price over 700 Years," unpublished. http://ideas.repec.org/p/nbr/nberwo/5132.html 

\bibitem{feller}  W. Feller, \textit{An Introduction to Probabilty Theory
and Its Applications}, 3rd Edition, Vol. I, John Wiley \& Sons, New York
(1968).


\bibitem{pareto}  V. Pareto, {\it Corso Di Economia Politica}, Vol II, 325-385,
Einaudi, Torino (1953).


\bibitem{stati}  Inland Revenue, National Statisics, UK (http://www.inlandrevenue.gov.uk/stats/). U.S. Census Bureau (http://www.census.gov).

\bibitem{souma22}  W. Souma, Fractals, \textbf{9} No. 4, 463-470 (2001).

\bibitem{dragulescu}  A. Dr\u{a}gulescu and V.M. Yakovenko, Physica A 
\textbf{299}, 213 (2001).

\bibitem{dragulesco2}  A. Dr\u{a}gulescu and V. M. Yakovenko,  Eur. Phys. J. \textbf{B20}, 585-589, (2001).



\bibitem{marx1846} K. Marx, {\it The poverty of Philosophy}, Crarles Kerr, Chicago (1846). Pag. 55: ``The natural price of labor is nothing but the minimum wage."

\bibitem{marx1857} K. Marx, {\it Grundrisse}, Penguin, Middlesex (1857). Pag. 817: ``For the time being, necessary labor supposed as
such; i.e. that the worker always obtains only the minimum of wages. This
supposition is necessary, of course, so as to establish the laws of profit
in so far as they are not determined by the rise and fall of wages or by
the influence of landed property. All these fixed suppositions themselves
become fluid in the further course of development."


\bibitem{marx1861} K. Marx, {\it Theories Of Surplus Value}, Parts I, II And III, Progress
Press, Moscow (1861). Part I, pag. 46: ``Since the value of raw and other materials is
given, while the value of the labour-power is equal to the minimum of
wages, this surplus-value can clearly only consist in the excess of labour
which the labourer returns to the capitalist over and above the quantity of
labour that he receives in his wage." Part II, pag. 223; Deriding Smith's discussion of wage
determination, Marx observes: ``In fact the chapter contains nothing
relevant to the question except the definition of the minimum wage, alias
the value of labour-power."


\bibitem{oxford}  J. Black, \textit{Dictionary of Economics}, Oxford
University Press Inc., New York (2002).

\bibitem{dues}  J.S. Duesenberry, \textit{Income, Saving, and the Theory of
Consumer Behavior}, Harvard University Press, Cambridge, MA (1949).


\bibitem{fn1}  It may be of interest to note that Pareto, in overt polemic
with some of his contemporaries, believed in the primacy of empirical data
over preconceived theories. He is also generally credited with having
introduced power laws as relevant phenomenological distributions.\cite
{pareto}

\bibitem{aoyaha}  H. Aoyama, W. Souma, Y. Nagahara, M.P. Okazaki, H.
Takayasu and M. Takayasu, Fractals \textbf{8}, 293 (2000); W.J. Reed,
Physica A (in press); A. Dragulescu and V.M. Yakovenko, Eur. Phys. J B 
\textbf{20}, 585 (2001); W. Souma, Fractals \textbf{9}, 463 (2001).

\bibitem{souma}  W. Souma, Y. Fujiwara and H. Aoyama, cond-mat/0108482
(2001).

\bibitem{levy}  M. Levy and S. Solomon, Physica A \textbf{242}, 90 (1997).



\bibitem{quandrini}  V. Quadrini and J.V. Rios-Rull, Fed. Res. Bank of
Minneapolis Quarterly Rev. \textbf{21}, 22 (1997) and references therein.

\bibitem{soto}  H. DeSoto, \textit{The Mystery of Capital}, Basic Books
(2000).

\bibitem{sands}  J.E. Sands, Wealth, Income and Intangibles, University of
Toronto Press (1963); R.R. Doane, T\textit{he Measurement of American Wealth}%
, Harper \& Brothers (1933).

\bibitem{west99}  B.J. West, \textit{Physiology, Promiscuity and Prophecy at
the Millennium: A Tale of Tails}, World Scientific (1999).

\bibitem{solomon}  S. Solomon and P. Richmond, Physica A \textbf{299}, 188
(2001); Z. Huang and S. Solomon, Physica A \textbf{306}, 412 (2002); O.
Malcai, O. Biham, P. Richmond and S. Solomon, Phys. Rev. E. \textbf{66},
031102 (2002).

\bibitem{abul}  A.Y. Abul-Magd, Phys. Rev. E \textbf{66}, 057104 (2002).

\bibitem{west90}  K. Lindenberg and B.J. West, \textit{The Nonequilibrium
Statistical Mechanics of Open and Closed Systems,} VCH Publishers, New York
(1990).

\bibitem{sornette}  D. Sornette, Phys. Rev. E \textbf{57}, 4811 (1998); M.
Levy and S. Solomon, Int. J. Mod. Phys. C \textbf{7}, 595 (1996).

\bibitem{ispolatov}  S. Ispolatov, P.L. Krapivsky and S. Redner, Eur. Phys.
J. B \textbf{2}, 267 (1998).

\bibitem{hayes}  B. Hayes, ``Follow the Money'', American Scientist, pg.
400, Sept-Oct (2002).


\bibitem{Chakraborti}  A. Chakraborti, B.K. Chakrabarti,  Eur. Phys. J.  \textbf{B17}, 167 (2000).

\bibitem{Chatterjee} A. Chatterjee, B. K. Chakrabarti and S. S. Manna, in press on Physica A (2004).


\bibitem{huang}  K. Huang, \textit{Statistical Mechanics}, Wiley, New
York(1987).

\bibitem{scafwest} N. Scafetta, B. J. West and S. Picozzi,  in press on a special issue of Physica D   to be entitled ``Anomalous Distributions, Nonlinear Dynamics, and Nonextensivity" (2003). This paper is part of a conference proceedings for the international Workshop on Anomalous Distributions, Nonlinear Dynamics and Nonextensivity, Nov 6-9 2002, Santa Fe (NM). The work was presented by N. Scafetta.



\bibitem{mauro}  M. Boianovsky and V.J. Tarascio, ``Mechanical inertia and
economic dynamics: Pareto on business cycles," J. of the History of Economic
Thought, \textbf{20} 5 (1998).

\bibitem{fn2}  Even in societies generally regarded as utterly impoverished,
wealth can still be relatively abundant. This paradox is identified and
discussed by De Soto\cite{soto}.



\bibitem{huang2}  Z. Huang and S. Solomon, Physica A \textbf{294}, 503
(2001).

\bibitem{watts}  D.J. Watts and S.H. Strogatz, Nature \textbf{393}, 440
(1998).


\bibitem{montroll2}  E.W. Montroll and M.F. Shlesinger, J. Stat. Phys. 
\textbf{32}, 209 (1983).


\bibitem{baekgaard}  H. Baekgaard, ``The Distribution of Household Wealth in
Australia'', NATSEM Discussion Paper No. 34, Sept. (1998).


\bibitem{pqretoelit} V. Pareto, ``The Rise and Fall of the Elites: An Application of
Theoretical Sociology."  Reprint, New Brunswick, New Jersey:
Transaction Books, (1991).

\bibitem{paretomarx} In his ``Treatise on
General Sociology'' (1935, reprint, New York: AMS Press, 1983) (par. 2462) V. Pareto criticized the socialist revolutionary expectations by stating: ``All revolutionaries
proclaim, in turn, that previous revolutions have ultimately ended up by
deceiving the people; it is their (socialist) revolution alone which is the true
revolution. ...  Unfortunately this true revolution, which is to bring men an unmixed
happiness, is only a deceptive mirage that never becomes a reality." 


\bibitem{pricedisc} J. M. Henderson and R. E. Quandt, {\it ``Microeconomic theory: a mathematical approach,"} McGraw-Hill, USA (1971).  The  {\it price discrimination in monopoly theory} says that a seller with a degree of monopoly power has the ability to price discriminate.  This means being able to charge a different price to different customers. In fact, for example, theaters often charge youngers or students less than others;  grocery stores have  lower prices for people who bother to check sale periods and look for clip coupons; some companies produce almost similar products but try to promote one as a prestige brand supposed for wealthy people at a much higher price; and so on.



\end{thebibliography}
\end{document}